\documentstyle[sprocl,epsfig]{article}

\begin{document}

\title{PHOTON COUNTING SAMPLING OF PHASE SPACE}
\author{KONRAD BANASZEK AND KRZYSZTOF W\'{O}DKIEWICZ}
\address{Instytut Fizyki Teoretycznej, Uniwersytet Warszawski,
Poland}

\maketitle

\abstracts{
The recently proposed scheme for direct sampling of the quantum
phase space by photon counting is discussed within the Wigner
function formalism.}

\noindent
First complete experimental characterisation of the quantum state of a
single light mode was demonstrated by Smithey {\em et
al.\/}\cite{Smith} In their work, numerical tomographic algorithms were
applied to reconstruct the Wigner function from homodyne statistics.
Recently, a novel scheme for measuring the quantum state of light by
photon counting has been proposed.\cite{Bana1} In contrast to
tomographic techniques, the newly proposed method allows one to measure
quasidistribution functions directly at a selected point of the phase
space.  In this contribution we show that the Wigner phase space
representation provides a clear and intuitive interpretation of this
scheme.

The principle of direct sampling of the quantum phase space by
photon counting is to measure the photon statistics $\{p_n\}$ of
the signal field $\hat{a}_{S}$, superposed on a probe field
$\hat{a}_{P}$ by means of a beam splitter with the power
transmission $T$. The measured photocount statistics is used to
calculate the alternating series $\sum_{n=0}^{\infty} (-1)^{n}
p_{n}$, which is given by the expectation value of the following
normally ordered operator expressed in terms of the signal and
the probe fields:
\begin{equation}
\hat{\Pi} = \; : \exp[-2(\sqrt{T} \hat{a}_{S}^{\dagger}  -
\sqrt{1-T} \hat{a}_{P}^{\dagger}) ( \sqrt{T}\hat{a}_{S} 
- \sqrt{1-T}\hat{a}_{P})]:.
\end{equation}

We will interpret this observable using the quantum mechanical
phase space formalism introduced by Eugene Wigner. For this
purpose we will represent $\hat{\Pi}$ as an integral of the
product of two quantities depending only on the signal or the
probe mode operators:
\begin{eqnarray}
\hat{\Pi} & = & \frac{2}{\pi} \int \mbox{\rm d}^{2} \beta
\;
:\exp [-2(\sqrt{T}\beta^{\ast} - \hat{a}^{\dagger}_{P})
(\sqrt{T}\beta - \hat{a}_{P})]: \nonumber \\
& & \times :\exp[-2(\sqrt{1-T}\beta^{\ast} - \hat{a}^{\dagger}_{S})
(\sqrt{1-T}\beta - \hat{a}_{S})] :.
\end{eqnarray}
As the $P$ mode is to be used as a probe for the $S$ mode, we
assume that both the modes are not correlated. Evaluating the
quantum expectation value of the above integral representation
yields the following formula:
\begin{equation}
\langle\hat{\Pi}\rangle = \frac{\pi}{2(1-T)} 
\int \mbox{\rm d}^{2} \beta \; W_{S}(\beta) W_{P}
(\sqrt{T/(1-T)}\beta),
\end{equation}
which shows that the measured quantity is simply the phase space
integral of the product of the signal and the probe Wigner
functions $W_{S}(\beta)$ and $W_{P}(\beta)$. Thus, the
quasidistribution function in the signal phase space is
``sampled'' by the probe Wigner function. Of course, the area of
the ``patch'' of the probe Wigner function must be at least that
imposed by the uncertainty principle.  The essential advantage
of the newly proposed scheme is that  the parameterisation of
the probe Wigner function in Eq.~(3) is rescaled by the factor
$\sqrt{T/(1-T)}$, which may take {\em any\/} positive value for
$0 < T < 1$. When $T > 1/2$, the probe Wigner function becomes
effectively ``squeezed'' in all directions, and consequently the
resolution of the phase space measurement goes beyond the
uncertainty relation limit. This feature is depicted
schematically in Fig.~1.  Let us stress that the rescaled Wigner
function does not appear physically in the setup; it is only a
tool for the phase space interpretation of the measured
observable.

\begin{figure}
\begin{minipage}[c]{7.8cm}
\setlength{\unitlength}{1.2cm}
\begin{picture}(6.5,4)
\put(0,0){\makebox(0,0)[lb]{\epsfig{file=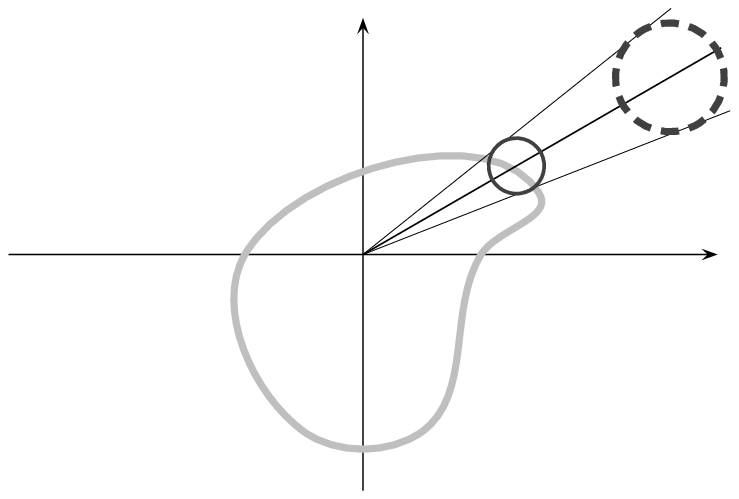}}}
\put(2,0.3){\makebox(0,0){\footnotesize $W_S(\beta)$}}
\put(6,2.7){\makebox(0,0){\footnotesize $W_P(\beta)$}}
\put(4,3.3){\makebox(0,0){\footnotesize $W_P(\sqrt{\frac{T}{1-T}}\beta)$}}
\end{picture}
\end{minipage}
\hfill
\begin{minipage}[c]{4cm}
\footnotesize
Figure 1:
Phase space interpretation of the observable 
$\hat{\Pi}$ measured in
the photon counting experiment.
\end{minipage}
\end{figure}

An important class of probe states which can be easily generated
in a laboratory are coherent states $|\alpha\rangle$. In
this particular case the outcome of the measurement can be
expressed using an $s=-(1-T)/T$ ordered quasidistribution
function of the signal mode:
\begin{equation}
\langle\hat{\Pi}\rangle 
= 
\frac{\pi}{2T} W_S \left(\sqrt{\frac{1-T}{T}} \alpha ;
-\frac{1-T}{T}\right).
\end{equation}
In the limit $T\rightarrow 1$ the ordering parameter approaches
zero, which corresponds to the direct measurement of the Wigner
function of the signal field. The Wigner function is determined
at a point defined by the amplitude and the phase of the probe
coherent state. By changing these two parameters, the complete
Wigner function can be scanned point--by--point, without using 
complex numerical algorithms.

This result is easily understood within the Wigner function
formalism. In the limit $T\rightarrow 1$ the rescaled Wigner
function of the coherent probe approaches the shape of the delta
function, and therefore the integral in Eq.~(3) picks up the
value of the signal Wigner function at a single phase space
point.  Additionally, the rescaling moves the probe Wigner
function towards the phase space origin, which is reflected by
the decreasing factor multiplying $\alpha$ in Eq.~(4).
Therefore, large intensity probe fields have to be used to scan
the required region of the phase space.

In a realistic setup, there is a limitation on the resolution of
the phase space sampling imposed by the non--unit quantum
efficiency $\eta$ of the photodetector. Furthermore, rigorous
statistical analysis shows that an attempt of numerical
compensation for this deleterious effect fails due to rapidly
exploding statistical error. We illustrate this with Fig.~2,
where Monte Carlo simulations of the photon counting experiment
for the one--photon Fock state are superimposed on the exact
expectation value with the analytically calculated statistical
error. Detailed discussion of these issues can be found in
Ref.~3. 

This work was partially supported by the Polish KBN grant 2~P03B~006~11.

\begin{figure}
\begin{center}
\setlength{\unitlength}{1cm}
\begin{picture}(11.5,4)
\put(0,0){\makebox(0,0)[lb]{\epsfig{file=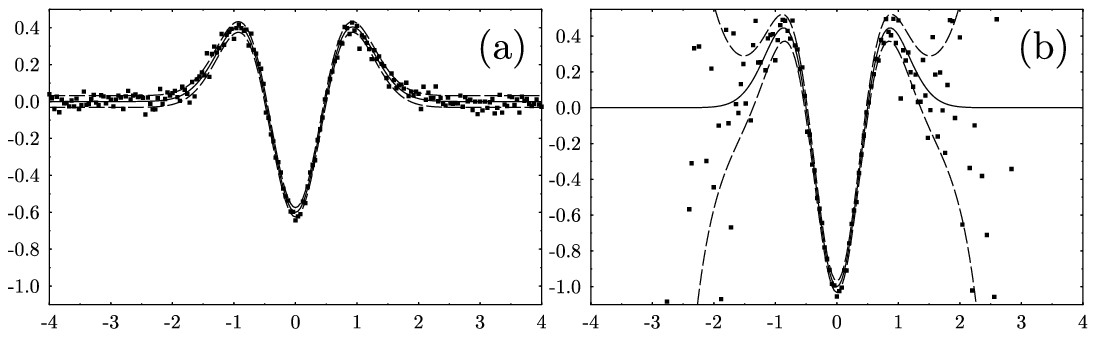}}}
\put(0.2,3.5){\makebox(0,0){\footnotesize$\langle\hat{\Pi}\rangle$}}
\put(3.5,0.3){\makebox(0,0){\footnotesize$\sqrt{(1-T)/T}\alpha$}}
\put(9.5,0.3){\makebox(0,0){\footnotesize$\sqrt{(1-T)/T}\alpha$}}
\end{picture}
\end{center}
\begin{minipage}{\textwidth}
\footnotesize
Figure 2: The one--photon Fock state quasidistribution function
reconstructed from $1000$ Monte Carlo events at each phase space
point (a) without and (b) with compensation for the non--unit
detector efficiency $\eta=80\%$, in the limit $T\rightarrow 1$.
\end{minipage}
\end{figure}

\end{document}